\documentclass[12pt]{article}
\usepackage[dvips]{graphicx}
\usepackage{pdproc}
\usepackage[dvips]{color}

  \makeatletter 
  \def\@cite#1{[#1]} 
  \makeatother    
\begin{document}

\renewcommand{\thefootnote}{\alph{footnote}}


\newcommand\lsim{\mathrel{\rlap{\lower4pt\hbox{\hskip1pt$\sim$}}
    \raise1pt\hbox{$<$}}}
\newcommand\gsim{\mathrel{\rlap{\lower4pt\hbox{\hskip1pt$\sim$}}
    \raise1pt\hbox{$>$}}}

\newcommand\cb{\textcolor{cyan}}
\newcommand\mhalf{M_{1/2}}

\newcommand\etal{{\it et\,al.}}

\title{
Flavor Structure of Scherk-Schwarz Supersymmetry Breaking and
 Constraints from Low Energy Processes
\footnote{Talk given by K.O. at the 12th International
Conference on Supersymmetry and Unification of Fundamental Interactions (SUSY04),
June 17-23, 2004, Epochal Tsukuba, Tsukuba}
}

\author{HIROYUKI ABE, KIWOON CHOI, KWANG-SIK JEONG AND KEN-ICHI OKUMURA}

\address{ 
Department of Physics, 
Korea Advanced Institute for Science and Technology, 
Daejeon, 305-701, Korea
\\ {\rm E-mail:abe@hep.kaist.ac.kr, kchoi@hep.kaist.ac.kr, ksjeong@hep.kaist.ac.kr, okumura@hep.kaist.ac.kr}}

\abstract{5D orbifold has two attractive features: 
quasi-localized matter fields can naturally reproduce hierarchical
 Yukawa coupling, while supersymmetry breaking is inherently built in
 by the Scherk-Schwarz mechanism. 
We examine the consequence of this quasi-localization and the Scherk-Schwarz
supersymmetry breaking in low energy flavor violating processes, under the assumption that physics below the compactification scale
 is described by the minimal supersymmetric standard model (MSSM)
and find $BR(\mu \to e,\gamma)$ and $\epsilon_K$ impose stringent
 constraint on flavor structure of the Scherk-Schwarz supersymmetry breaking.
Chirality measurement in lepton flavor violating processes
 is crucial to deduce the surviving flavor structure.
}
\normalsize\baselineskip=15pt
\vspace{1em}

Localization of matter fields along the extra-dimension
 can naturally  reproduce observed Yukawa
 hierarchy \cite{paper1}, while the Scherk-Schwarz mechanism
 provides a simple alternative way to break supersymmetry \cite{paper2}.  
5D orbifold $S^1/Z_2$ is a minimal realization of
 both of these attractive features of introducing new dimensions.
Let us first define the model using 5D bulk action on $S^1$
 in terms of 4D superfield representation \cite{paper3}, 
\begin{eqnarray}
S_{bulk} &=&
\int d^4x \int dy \,\left[\,
\int d^4 \theta \,
\left\{
Re(T)\left( \Phi_I \Phi^*_I+\Phi^c_I
				  \Phi^{c*}_I\right)
+\frac{1}{g_{5 a}^2}
\frac{1}{Re(T)}
\left(
\partial_5 V -\sqrt{2}Re(\chi)
\right)^2
\right\}
\right.
\nonumber\\
&&
\left.\phantom{\int d^5x \,[}
+ \left(\,\int d^2\theta\,
\Phi^c_I\left(\partial_y-\frac{1}{\sqrt{2}}\chi+M_I T \right)\Phi_I
+\frac{1}{4g^2_{5a}}T W^{a\alpha} W^a_{\alpha} + {\rm h.c.}\right)
\,\right],
\label{eq1}
\end{eqnarray}
where 4D chiral multiplets $(\Phi_I, \Phi^c_I)$ and 4D vector and chiral
 multiplets $(V, \chi)$ constitute 5D hyper and vector
 multiplets respectively. 5D supersymmetry forbids bulk Yukawa coupling.
Here, radius modulus R of $S^1$, satisfying $x^5 = R y~~(0\le y<2 \pi)$,
 is promoted to the radion chiral multiplet $T$ so that 4D N=1 supersymmetry is
 manifest. In 5D supergravity, it is identified as
 a part of 5D supergravity multiplet \cite{paper2}.
Because we are interested in 4D chiral theories, we orbifold $S_1$ by $Z_2$
imposing boundary
 conditions,
\begin{equation}
{V}^a(-y)={V}^a(y),\quad
{\chi}^a(-y)=-{\chi}^a(y),\quad
\Phi_I (-y) =\Phi_I (y),\quad
\Phi_I^c(-y) = -\Phi^c_I (y),
\label{eq2}
\end{equation}
 so that 4D $N=2$ supersymmetry is broken to $N=1$.
After orbifolding, the surviving bulk mass should be regarded as
a kink mass, $M_I \rightarrow M_I\epsilon(y)$ respecting $Z_2$.
Then the zero mode equation for $\Phi_I$ fermion component, $\psi_I=
\chi(x)\tilde{\phi}_{0I}(y)$ is given by
$
\big(\,\partial_y+M_IT\epsilon(y)\,\big)\tilde{\phi}_{0I}=0,
$
yielding the zero mode wavefunction
$
\tilde{\phi}_{0I}\propto e^{-M_IT|y|},
$
which shows that it is quasi-localized at $y=0$ if $M_I>0$,
and at $y=\pi$ if $M_I<0$.
Inspired by this observation, we assume Higgs fields
 exist on the fixed point at $y=0$ and
introduce the following brane action,
\begin{equation}
S_{brane}=\int d^4 x \int d^2 \theta \, H
\int d y
\,\delta(y) 
\lambda_{IJ}^{(5)} \Phi_I \Phi_J  
+ {\rm h.c}\,,
\label{eq3}
\end{equation}
After redefinition of the chiral fields,
${\Phi}_I \,\rightarrow\, e^{M_I T |y|} \Phi_I$,~ 
${\Phi}^c_I \,\rightarrow\, e^{-M_I T |y|} \Phi^c_I$, 
and appropriate renormalization,
we can extract the 4D effective action for $\Phi_I$ zero mode,
\begin{eqnarray}
&&S_{4D} = \int d^4x
\left[\int d^4\theta
\left(
\Phi_I \Phi_I^\ast+H H^\ast
\right)
+\left( \int d^2\theta\, y_{IJ} H \Phi_I \Phi_J
+{\rm H.c.}
\right)
\right],
\label{eq4}
\end{eqnarray}
where we use the same symbol $\Phi_I$ for the corresponding canonical zero-modes and,
\begin{equation}
y_{IJ}=\lambda_{IJ}^{(4)}
\sqrt{\frac{N_IN_J}{(\epsilon^{-2N_I}-1)(
\epsilon^{-2N_J}-1)}}
,~~{\rm for}~~
\lambda^{(4)}_{IJ}=\frac{\ln(1/\epsilon)}{\pi R}\lambda^{(5)}_{IJ}
,~~
N_I=-\frac{\pi R}{
\ln(1/\epsilon)}M_I.
\label{eq5}
\end{equation}
Here, we choose
$
\epsilon\,\equiv \,\mbox{Cabibbo angle}\,
\approx 0.2
$
 for later convenience. 
Naive dimensional argument shows
$\lambda^{(4)}_{IJ}\sim {\cal O}(1)$  while
hierarchical $y_{IJ}$ is naturally realized for $N_I>0$.

So far the 4D effective theory is supersymmetric, however,
the remaining supersymmetry can be  generally broken
 by non-vanishing radion F-term, $F^T$ \cite{paper4}. 
The resultant soft terms are easily obtained by standard calculus as follows,
\begin{equation}
S_{soft} = -\int d^4x \, \Big(\,
\frac{1}{2}m^2_{IJ}\phi^*_I\phi_J+
A_{IJ}\phi_I\phi_J h+
\frac{1}{2}M_a\lambda^a\lambda^a+{\rm h.c.}\,\Big),
\label{eq6}
\end{equation}
\begin{eqnarray}
&&
M_a= M_{1/2}=-\frac{F^T}{2R}\,,~~~~~
A_{IJ}=2y_{IJ}\ln(1/\epsilon)\frac{F^T}{2R}\left(\,
\frac{N_I}{1-\epsilon^{2N_I}}+
\frac{N_J}{1-\epsilon^{2N_J}}\,\right),
\nonumber \\
&&
m^2_{I\bar{J}}=\delta_{IJ}\left(\,
2\ln(1/\epsilon)\frac{N_I}{\epsilon^{N_I}-
\epsilon^{-N_I}}\frac{F^T}{2R}\,\right)^2,
\label{eq7}
\end{eqnarray}
where lower-case fields denote scalar components of the chiral multiplets
 and $\lambda^a$ represents gaugino.
In $F^T/R<<1$ limit, 
this mass spectrum exactly matches with that of 
the Scherk-Schwarz supersymmetry
breaking (SS SUSY breaking), where twisted boundary condition is imposed
 on $SU(2)_R$ over $S^1/Z_2$ \cite{paper5}.
It is highly dependent on the flavor structure given in
 (\ref{eq5}), therefore becomes a source of
 dangerous flavor changing processes at the electro-weak scale. 
In the following, we present a brief summary of systematic analysis
 of this low energy constraint on the flavor structure of the SS
 SUSY breaking.

The above Yukawa couplings are quite similar to
the Yukawa couplings in Frogatt-Nielsen models
with $N_I$ being identified as the $U(1)_F$ charges.
More explicitly,
 assuming that mass spectrum below the compactification scale is given by
 that of the minimal supersymmetric standard model (MSSM),
\begin{eqnarray}
y^u_{ij}\,\simeq \,
\lambda^{u\,(4)}_{ij}\epsilon^{X^u_i+X^q_j}\,,~~
y^d_{ij}\,\simeq \,
\lambda^{d(4)}_{ij}\epsilon^{X^d_i+X^q_j}\,,~~
y^\ell_{ij}\,\simeq \,
\lambda^{\ell(4)}_{ij}\epsilon^{X^e_i+X^\ell_j}\,,
\label{eq8}
\end{eqnarray}
where $u_i$, $d_i$ ($e_i$) denote the
 $SU(2)_L$ singlet quarks (leptons) 
 and $q_i$  ($\ell_i$) represents the doublet quarks (leptons).
 The effective flavor charge $X_I$ is given by
 $X_I = N_I~(X_I = 0 )$ for $N_I \ge 0~ (N_I\le 0)$.
The above soft terms also can be well approximated by $X_I$
and $N_I$,
\begin{eqnarray}
&&
A^u_{ij} \simeq  M_0 (X^u_i+X^q_j)\,y^u_{ij},~~
A^d_{ij} \simeq  M_0 (X^d_i+X^q_j)\,y^d_{ij},~~
A^\ell_{ij} \simeq M_0(X^e_i+X^\ell_j)\,y^\ell_{ij},
\nonumber \\
&&
m^{2(\tilde{\psi})}_{i\bar{j}} \simeq
\delta_{i\bar{j}}
M_0^2 \left\{
\begin{array}{ll}
|N^\psi_i|^2 \epsilon^{2|N^\psi_i|}&~(\,N^{\psi}_i \neq 0\,)\\
1/[2\ln(1/\epsilon)]^2&~(\,N^{\psi}_i = 0\,)
\end{array}
\right.
~~ {\rm for}~~~ M_0=-2\mhalf\ln (1/\epsilon).
\label{sssoftmssm}
\end{eqnarray}

To discuss the low energy observables, it is convenient to move from the
 above definition basis to the SCKM
 basis where quarks and leptons have diagonal mass matrices, {\it e.g.}
$
y^\ell_{ij}
 \longrightarrow (\overline{V}_e)_{ik} y^\ell_{k\ell}
 (\overline{V}_\ell^\dag)_{\ell j} = \hat{y^\ell}_i \delta_{ij}. 
$
In which the structure of the Yukawa coupling given in (\ref{eq8})
 is well reproduced by imposing constraints on the 
 unitary matrices,
$
|(\overline{V}_{e,\ell})_{ij}| \lsim  \epsilon^{|X^{e,\ell}_i-X^{e,\ell}_j|}
$
for given diagonal Yukawa couplings, $\hat{y^\ell}_i \sim {\cal O}(\epsilon^{X^e_i+X^\ell_i})$.
Similar discussion is applied for quarks.
These constraints are directly translated into the probable magnitudes
 of mass-insertion parameters at the electro-weak scale defined like,
 $(\delta^\ell_{RL})_{ij}\equiv A^\ell_{ij}
 v_d/\sqrt{m^{2(\widetilde{e})}_{ii} m^{2(\widetilde{\ell})}_{jj}}$ 
, $(\delta^d_{RR}) \equiv 
 m^{2(\widetilde{d})}_{ij}/\sqrt{m^{2(\widetilde{d})}_{ii}m^{2(\widetilde{d})}_{jj}}$
and $(\delta^d_{LL}) \equiv 
 m^{2(\widetilde{q})}_{ij}/\sqrt{m^{2(\widetilde{q})}_{ii}m^{2(\widetilde{q})}_{jj}}$ in the SCKM basis \cite{paper5}. 
Note that if some of the effective charges are degenerate, unitarity
 relation like
$
(\overline{V}_e)_{ik} X^e_k (\overline{V}_e^\dag)_{kj} =
 (\overline{V}_e)_{i1} (\overline{V}_e^*)_{j1} (X^e_1-X^e_2)
-(\overline{V}_e)_{i3} (\overline{V}_e^*)_{j3} (X^e_2-X^e_3)
+\delta_{ij} X^e_2  \nonumber\\
$
can dramatically suppresses the mixing from the naive estimation.
This suppression mechanism is quite natural if underlying physics
quantizes the kink
 masses in some unit, which originate from
 $U(1)_{FI}$ or graviphoton charges in the supergravity formulation
 \cite{paper6}.
\vspace{-1em}
\begin{table}[h]
\begin{center}
\caption{ Lepton mass hierarchy vs constraint from $\mu \to e\gamma$.
$(\Delta^\ell_{RL(LR)})_{12}\equiv (\delta^\ell_{RL(LR)})_{12}/4.8\times 10^{-6}$.
 \label{table1}}
\footnotesize
\begin{tabular}{|c|c|c|c|c|c|c|c|}
\hline
\multicolumn{4}{|c|}{$\hat{y}^\ell_i/\hat{y}^\ell_3={\cal O}(\epsilon^5,\epsilon,1)$}&
\multicolumn{4}{|c|}{$\hat{y}^\ell_i/\hat{y}^\ell_3={\cal O}(\epsilon^6,\epsilon,1)$}\\
\hline
$X^e_{i}-X^e_3$ & $X^\ell_{i}-X^\ell_3$ &
$|(\Delta^\ell_{RL})_{12}|$ &
$|(\Delta^\ell_{LR})_{12}|$ &
$X^e_{i}-X^e_3$ & $X^\ell_{i}-X^\ell_3$ &
$|(\Delta^\ell_{RL})_{12}|$ &
$|(\Delta^\ell_{LR})_{12}|$ 
\\
\hline
$(5,1,0)$&$(0,0,0)$ & 3.2& 0.015
                   & $(6,1,0)$&$(0,0,0)$ & 0.80 & 0.040\\
$(0,0,0)$&$(5,1,0)$ & 0.015 & 3.2 
                    & $(0,0,0)$&$(6,1,0)$ & 0.040& 0.80\\
\hline
\hline
\multicolumn{4}{|c|}{$\hat{y}^\ell_i/\hat{y}^\ell_3={\cal O}(\epsilon^5,\epsilon^2,1)$}&
\multicolumn{4}{|c|}{$\hat{y}^\ell_i/\hat{y}^\ell_3={\cal O}(\epsilon^6,\epsilon^2,1)$}\\
\hline
$X^e_{i}-X^e_3$ & $X^\ell_{i}-X^\ell_3$ &
$|(\Delta^\ell_{RL})_{12}|$ &
$|(\Delta^\ell_{LR})_{12}|$ &
$X^e_{i}-X^e_3$ & $X^\ell_{i}-X^\ell_3$ &
$|(\Delta^\ell_{RL})_{12}|$ &
$|(\Delta^\ell_{LR})_{12}|$ 
\\
\hline
\multicolumn{4}{|c|}{}
                   &$(6,2,0)$&$(0,0,0)$& 3.2 & 0.60\\ 
\multicolumn{4}{|c|}{ No surviving models}
                   &$(2,2,0)$&$(4,0,0)$& 1.6& 3.2\\
\multicolumn{4}{|c|}{ with mild tuning.}
                   &$(4,0,0)$&$(2,2,0)$& 3.2 & 1.6\\  
\multicolumn{4}{|c|}{}
                   &$(0,0,0)$&$(6,2,0)$& 0.60 &3.2\\
\hline
\end{tabular}
\end{center}
\end{table}

We have explored various lepton flavor violating (LFV) and FCNC
 processes
 and find $\mu \to e,\gamma$ and $\epsilon_K$ give the most
 stringent constraint on the flavor structure of the SS SUSY breaking.
Table \ref{table1} lists surviving set of $X^{e,\ell}_i$ from $\mu \to
 e,\gamma$ with $(\delta^\ell_{RL(LR)})_{12}$ divided by a
 value which saturates the current upper-bound,
 $BR(\mu \to e,\gamma) = 1.2 \times 10^{-11}$ for $|M_{1/2}|=500$ GeV.
The table shows that at least either $X^e$ or $X^\ell$ should
 have degenerate charges to satisfy the bound. 
Table \ref{table2},\ref{table3} show typical predictions of the models including other LFV processes.
Note that degenerate charges of $\ell$ ($e$) yields the chirality of the
 processes opposite (similar) to the seesaw induced neutrino mass models \cite{paper7}.
%
\begin{table}[h]
\begin{center}
\vspace{-1em}
\caption{Predictions of LFV rates for
 $y^\ell_i={\cal O}(\epsilon^8,\epsilon^3,\epsilon^2)$ and $|\mhalf|=500
 {\rm GeV}$. $t_\beta\equiv \tan\beta$.\label{table2}}
\footnotesize
\begin{tabular}{|c|c|c|c|c|}
\hline
$N^e_i$ & $N^\ell_i$ &
$BR(\mu^+_R \to e^+_L,\gamma)$ &
$BR(\tau^+_R \to e^+_L,\gamma)$ &
$BR(\tau^+_R \to \mu^+_L,\gamma)$
\\
\hline
$6,1,-2$&$2,2,2$
 &$9.2(1+0.11t_\beta)^2\times 10^{-12}$
 &$-$
 &$2.4(1+0.091t_\beta)^2\times 10^{-8}$\\
$8,3,2$&$-1,-1,-1$
 &$7.8\times 10^{-12}$
 &$-$
 &$2.4(1+0.091t_\beta)^2\times 10^{-8}$\\
\hline
$N^e_i$ & $N^\ell_i$ &
$BR(\mu^+_L \to e^+_R,\gamma)$ &
$BR(\tau^+_L \to e^+_R,\gamma)$ &
$BR(\tau^+_L \to \mu^+_R,\gamma)$
\\
\hline
$-1,-1,-1$&$8,3,2$
 &$7.7\times 10^{-12}$
 &$-$
 &$2.1(1+0.060t_\beta)^2\times 10^{-8}$\\
$2,2,2$&$6,1,-2$
 &$7.7(1+0.075t_\beta)^2\times 10^{-12}$
 &$-$
 &$2.1(1+0.060t_\beta)^2\times 10^{-8}$\\
\hline
\end{tabular}
\end{center}
\end{table}
\begin{table}[h]
\begin{center}
\vspace{-1em}
\caption{Same as table 2 for $y^\ell_i={\cal O}(\epsilon^8,\epsilon^4,\epsilon^2)$.
\label{table3}}
\footnotesize
\vspace{-0.8em}
\begin{tabular}{|c|c|c|c|c|}
\hline
$N^e_i$ & $N^\ell_i$ &
$BR(\mu^+_R \to e^+_L,\gamma)$ &
$BR(\tau^+_R \to e^+_L,\gamma)$ &
$BR(\tau^+_R \to \mu^+_L,\gamma)$
\\
\hline
$2,2,-2$&$6,2,2$
 &$3.1\times 10^{-11}$
 &$3.4\times 10^{-9}$
 &$3.3 \times 10^{-9}$\\
$4,4,2$&$4,-2,-2$
 &$3.4(1+0.053t_\beta)^2\times 10^{-11}$
 &$3.7(1+0.047t_\beta)^2 \times 10^{-9}$
 &$3.6(1+0.047t_\beta)^2 \times 10^{-9}$\\
\hline
$N^e_i$ & $N^\ell_i$ &
$BR(\mu^+_L \to e^+_R,\gamma)$ &
$BR(\tau^+_L \to e^+_R,\gamma)$ &
$BR(\tau^+_L \to \mu^+_R,\gamma)$
\\
\hline
$4,-2,-2$&$4,4,2$
 &$3.1(1+0.032 t_\beta)^2\times 10^{-11}$
 &$3.4(1+0.030t_\beta)^2 \times 10^{-9}$
 &$3.3(1+0.030 t_\beta)^2 \times 10^{-9}$ \\
$6,2,2$&$2,2,-2$
 &$3.1\times 10^{-11}$
 &$3.4\times 10^{-9}$
 &$3.3\times 10^{-9}$ \\
\hline
\end{tabular}
\end{center}
\end{table}
\begin{table}[h!]
\begin{center}
\vspace{-1em}
\caption{Quark mass hierarchy vs $\epsilon_K$.
\label{table4}}
\footnotesize
\vspace{-0.8em}
\begin{tabular}{|c|c|c|c|c|}
\hline
$N^q_i$ & $N^d_i$ &
$\Im(\delta^d_{LL})^2_{12}/(1.5\times 10^{-2})^2$ &
$\Im(\delta^d_{RR})^2_{12}/(1.5\times 10^{-2})^2$ &
$\Im[(\delta^d_{RR})_{12}(\delta^d_{LL})_{12}]/(2.2\times 10^{-4})^2$ \\
\hline
$3,2,-1$ & $3,2,2$ &$1.5\times 10^{-2}$ &$7.1\times 10^{-1}$ & $466$\\
$3,2,-1$ & $4,2,2$ &$1.5\times 10^{-2}$ &$2.8\times 10^{-2}$ & $93$\\
\hline
$3,2,-1$ & $3,3,2$ &$1.5\times 10^{-2}$ &$2.8\times 10^{-2}$ & $93$\\
$3,2,-1$ & $4,3,2$ &$1.5\times 10^{-2}$ &$5.7\times 10^{-3}$ & $42$\\
\hline
\end{tabular}
\end{center}
\end{table}
On the other hand, table \ref{table4} shows the constraints from the CP violating
parameter of $\overline{K}^0$-$K^0$ mixing, $\epsilon_K$, where the
mass-insertion parameters are divided by values saturating the observation
$\epsilon_K=2.282 \times 10^{-3}$ for $|M_{1/2}|=500$ GeV.
Models in the table
 can reproduce the quark masses and CKM
angles with moderate tuning of the ${\cal O}(1)$ parameters, however, the
 $\epsilon_K$ constraint requires further ${\cal O}(10^{1-2})$ fine-tuning.
However,
if we allow abnormally
large or small ${\cal O}(1)$ parameters by one order of $\epsilon \approx 0.2$,
 we can eliminate dangerous right-handed
mixing by choosing degenerate $N^d_i$ and safely satisfy the constraint. 
We have also examined $\Delta M_K$, $\Delta M_{B_{d,s}}$, $\epsilon'/\epsilon$
, $b \to s,\gamma$, and CP asymmetry in $B_d \to \phi K_S$,
 however, can not find any meaningful constraints or
 predictions beyond the standard model.
%
\section{Acknowledgments}
\vspace{-.5em}
This work is supported by KRF PBRG 2002-070-C00022.
\vspace{-.5em}
\bibliographystyle{plain}

\end{document}